\documentclass[11pt]{iopart}
\usepackage{verbatim}
\usepackage{lineno,hyperref}
\modulolinenumbers[5]
\usepackage{color}
\usepackage[]{subfigure}
\usepackage{placeins} 
\usepackage[english]{babel}  
\usepackage[utf8x]{inputenc}
\usepackage{graphicx}
\usepackage{booktabs}
\usepackage{tabularx}
\usepackage{multirow}

\usepackage{geometry}
\begin{document}

\title[In search of the ground state crystal structure of Ta$_2$O$_5$]{In search of the ground state crystal structure of Ta$_2$O$_5$ from ab-initio and Monte Carlo simulations}

\author{Andrea Pedrielli\textsuperscript{1, 2}, Nicola Maria Pugno \textsuperscript{2, 3}, Maurizio Dapor \textsuperscript{1,4},  Simone Taioli \textsuperscript{1,4, *}}

\address{\textsuperscript{1} European Centre for Theoretical Studies in Nuclear Physics and Related Areas (ECT*) - Fondazione Bruno Kessler, Trento, Italy} 
\address{\textsuperscript{2} Laboratory for Bioinspired, Bionic, Nano, Meta Materials \& Mechanics, Department of Civil, Environmental and Mechanical Engineering, University of Trento, 38123 Trento, Italy}
\address{\textsuperscript{3} School of Engineering and Materials Science, Queen Mary University of London, Mile End Road, London E1 4NS, United Kingdom}
\address{\textsuperscript{4} Trento Institute for Fundamental Physics and Applications (TIFPA-INFN), Trento, Italy}

\address{\textsuperscript{*}Corresponding author at: European Centre for Theoretical Studies in Nuclear Physics and Related Areas (ECT*-FBK) (S. Taioli).}
\ead{taioli@ectstar.eu}

\small

\begin{abstract}
Tantalum oxides (Ta$_2$O$_5$) are characterised by attractive physical and chemical properties, such as high dielectric constants and anti-reflection behaviour. Recently, Ta$_2$O$_5$ nanoparticles have also been proposed as possible enhancers of the relative biological effectiveness in hadrontherapy for cancer treatment.
In principle, their electronic properties can be accurately investigated from first-principles simulations. However, the existence of several stable polymorphs of these oxides represents a major difficulty in order to calculate and disentangle their respective spectral features. To assess this problem, we use linear-response time-dependent density functional to investigate the energy loss function $\mbox{Im} (-1/\bar{\epsilon})$, which is a unique fingerprint of the material, in the optical limit for various polymorphs. We show that the experimental reflection energy loss signals can be rationalized and interpreted by assuming that the $\gamma$-phase of Ta$_2$O$_5$ represents the underlying structural model. We notice that both the inclusion of local field effects and spin-orbit coupling are crucial to compute the energy loss functions of this material. Finally, to further validate the $\gamma$-Ta$_2$O$_5$ polymorph as a model for experimental tantalum oxide, we compute the reflection energy loss spectra using a Monte Carlo approach, finding an excellent agreement with the experimental data.
\end{abstract}

\maketitle


\section{Introduction}
Tantalum pentoxide (Ta$_2$O$_5$) is a transition-metal oxide with a broad range of attractive characteristics such as high-dielectric and anti-reflection properties as well as wear resistance, thermal and chemical stability. Recently, Ta$_2$O$_5$ nanoparticles (NPs) have been proposed also as possible enhancers of the relative biological effectiveness (RBE) in hadrontherapy for cancer treatment \cite{McKinnon2016}. \\
\indent In all these applications the accurate assessment of the electronic excitation spectra is a key factor, which can be accomplished by using ab initio simulations \cite{Pedrielli2021} or experiments \cite{Fadanelli2015}.
Unfortunately, a long standing problem of bulk Ta$_2$O$_5$ is to be characterized by polymorphism \cite{ASKELJUNG2003250}. It can be found indeed in both amorphous and polycristalline arrangements, whereby several crystal structure co-exists, such as $\beta$-Ta$_2$O$_5$ \cite{Oehrlein1984}, $\beta'$-Ta$_2$O$_5$ \cite{HOLLERWEGER2015}, $\beta_{\mathrm{R}}$-Ta$_2$O$_5$ \cite{Ramprasad2003}, $\delta$-Ta$_2$O$_5$ \cite{Hiratani2002} and $\lambda$-Ta$_2$O$_5$ \cite{Lee2013}. Thus, the identification of a structural model of the most stable configuration of Ta$_2$O$_5$ to which the relevant electronic and optical properties obtained by experimental measurements are assigned is cumbersome. In particular, it is still debated how to relate the Ta$_2$O$_5$ polymorphs to the experimental structural information \cite{Lee2014}. 
Recently, a further low-energy structure -- the so-called $\gamma$-Ta$_2$O$_5$ \cite{Yang2018, Yuan2019} -- has been proposed. However, the reliability of this structural model to interpret the experimental spectral data is still to be thoroughly investigated. 

In order to find such model, in this work we reckon from first-principles the energy loss functions (ELF) $\mbox{Im} (-1/\bar{\epsilon})$, which represents the fingerprint of the material under investigation, of several Ta$_2$O$_5$ polymorphs and we compare them with ELF data derived from experimental reflection energy loss spectra (REELS) of the thermally grown oxide \cite{Fadanelli2015}. 
In particular, we show by using time-dependent density functional theory (TDDFT) that the $\gamma$-Ta$_2$O$_5$ polymorph is the only arrangement that captures all the features of the experimental excitation spectrum, confirming that this crystalline form may represent the most reliable model of the most stable tantalum pentoxide grown via thermal oxidation. 
Furthermore, we demonstrate that the inclusion of both local field effects (LFE), which are crucial in the assessment of the reflectivity and absorbance in materials with strong spatial inhomogeneity, and spin-orbit (SO) coupling is necessary to accurately compute the ELF of these materials. 

Finally, to compare directly the computed spectrum to the measured energy loss lineshape \cite{Fadanelli2015}, avoiding background subtraction \cite{Tougaard2004}, which is a procedure not free from uncertainty, we further check our findings by computing the REEL spectrum using Monte Carlo (MC) simulations. To carry out these calculations, 
we need to assess the elastic and inelastic scattering cross section of $\gamma$-Ta$_2$O$_5$. 

\section{Materials and methods}
\subsection{Materials structure}

\begin{figure}[htbp!]
\centering
\includegraphics[width=0.15\textwidth]{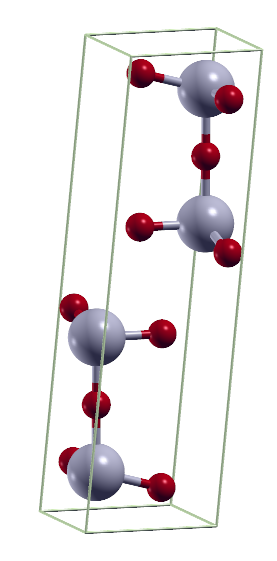}
\includegraphics[width=0.2\textwidth]{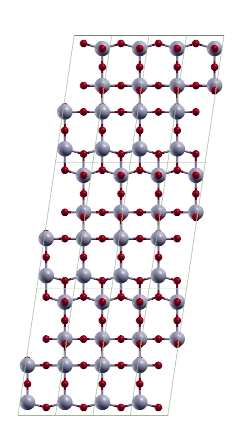}
\caption{Unit cell of the $\gamma$-Ta$_2$O$_5$ polymorph (left) and the corresponding 4$\times$3 supercell structure (right). The oxygen atoms are rendered in red, the tantalum atoms in grey.}
\label{fig:structureGAMMA}
\end{figure}

The following polymorphs were investigated in this work: $\beta$-Ta$_2$O$_5$ \cite{Oehrlein1984}, $\beta'$-Ta$_2$O$_5$ \cite{HOLLERWEGER2015}, $\beta_{\mathrm{R}}$-Ta$_2$O$_5$ \cite{Ramprasad2003}, $\delta$-Ta$_2$O$_5$ \cite{Hiratani2002}, $\lambda$-Ta$_2$O$_5$ \cite{Lee2013}, and $\gamma$-Ta$_2$O$_5$ \cite{Lee2014}. In our calculations we used the atomic coordinates reported therein. In particular, the unit cell of the $\gamma$-Ta$_2$O$_5$ polymorph is plotted in Fig. \ref{fig:structureGAMMA}. This crystal has a I41/amd symmetry, that is a tetragonal structure, which can be seen more clearly in the 4$\times$3 supercell reported in the right panel of Fig. \ref{fig:structureGAMMA}. Atomic units are used throughout. 

\subsection{Dielectric function}

The materials microscopic dielectric function can be reckoned by knowing the bare Coulomb potential $v_{\rm C}$ and the polarization function $\chi({\bf q},W)$ through \cite{RevModPhys.74.601, Weissker2010}:
\begin{equation}
\epsilon({\bf q},W)=1-v_{\rm C}({\bf q})\chi({\bf q},W) \, \mbox{.}
\end{equation}
The latter can be in turn obtained by solving the following Dyson--like equation:
\begin{equation}\label{Dyson}
    \chi^{-1}({\bf q},W)=\chi^{-1}_{0}({\bf q},W)-v_{\rm C}({\bf q})- f_{\rm xc}({\bf q},W) \, \mbox{,}
\end{equation}
where $\chi^{-1}_{0}({\bf q},W)$ is the non-interacting (or independent-particle) polarization obtained from the Kohn-Sham wavefunctions and $f_{\rm xc}({\bf q},W)$ is the energy and momentum dependent TDDFT kernel.
In this regard, we have used the adiabatic local density approximation (ALDA) kernel, which is related to the LDA exchange-correlation functional $v_{\rm  xc}[\rho]$ by:
\begin{equation}
    f_{\rm xc}(\textbf{r},t)= \left\{ \frac{d}{d\rho}v_{\rm  xc}[\rho]\right\}_{\rho=\rho(\textbf{r},t)},
\end{equation}
where $\rho$ is the DFT ground state density.

For periodic crystals one can exploit the translational symmetry, which allows one to write conveniently the microscopic dielectric function in the reciprocal space as a matrix, i.e. $\epsilon_{\mathbf{G},\mathbf{G'}}(\mathbf{q}, W) = \epsilon(\mathbf{q}+\mathbf{G}, \mathbf{q}+\mathbf{G'}, W)$, where $\mathbf{G}$ and $\mathbf{G'}$ are reciprocal lattice vectors, and $\mathbf{q}$ is the transferred momentum vector in the first Brillouin zone (1BZ).

The measured macroscopic dielectric function $\bar{\epsilon}( \mathbf{q}, W)$ and the computed microscopic function $\epsilon_{\mathbf{G},\mathbf{G'}}(\mathbf{q}, W)$ are related by the following expression \cite{Wiser}:
\begin{equation}\label{eq:eM}
\bar{\epsilon}( \mathbf{q}, W) =  \left [ \epsilon^{-1}_{\mathbf{G}=0,\mathbf{G'}=0}(\mathbf{q}, W)  \right ]^{-1}.
\end{equation}
LFE were included here by inverting the full dielectric matrix and then taking the head element of the inverse matrix.

\subsection{Energy loss function}

To describe the propagation of charged particles through matter using the dielectric formalism  \cite{Ritchie1959}, one relies on the ELF of the material, which is related to the macroscopic dielectric function as follows:
\begin{equation}\label{ELF}
\textrm{ELF} = \textrm{Im} \left [\frac{-1}{\bar{\epsilon}(\mathbf{q}, W)} \right ].
\end{equation} 

Here, we limit the ab initio calculation of the ELF to the optical limit (${\bf q} \rightarrow 0)$, as experimental optical data are the only available. The dielectric response function in the optical limit has been obtained using the linear response TDDFT implementation of the ELK code suite \cite{ELK} on top of the DFT ground state. ELK uses an all-electron Full-Potential Linearized Augmented-Plane-Wave (FP-LAPW) approach. 

The calculations were carried out in spin-polarized mode to include SO coupling. The local spin density approximation (LSDA) exchange correlation functional \cite{Perdew1992} has been used for the ground state calculations alongside the ALDA approximation for the time-dependent exchange correlation kernel. 

A 4 $\times$ 4 $\times$ 2 $k$-point grid for $\gamma$-Ta$_2$O$_5$ and $\beta$-Ta$_2$O$_5$ and a 4 $\times$ 2 $\times$ 2 $k$-point grid for the other polymorphs have been used. A number of empty bands equal to $60$ for each atom have been employed to obtain converged results up to $80$~eV. The ELF spectra were finally averaged over the three components of the polarization vector of the external electromagnetic field. 

\subsection{Extension of the energy loss function}

To run our Monte Carlo method for modelling the energy loss spectra of tantalum pentoxides, we need first to determine the inelastic and elastic scattering cross sections. The former can be obtained by knowing the dependence of the ELF on the entire spectrum of excitation energies $W$ and momentum transfers ${\mathbf{q}}$ \cite{Pedrielli2021,azzolini2017monte,azzolini2018anisotropic,taioli2020relative,azzolini2020comparison}.
Ab-initio calculations of the ELF over a large energy range are prohibitive owing to the high computational costs of including many electronic transitions to the excited states. Thus, while first-principles can be used for computing the ELF below 100 eV, its extension along the excitation energy axis has been performed by using the experimental NIST X-ray atomic data up to 30 keV \cite{NIST}. The ab-initio and experimental ELFs in the optical limit were matched smoothly and were fitted by Drude-Lorentz functions coupled with a momentum dependent broadening $\gamma_i$, which takes into account the momentum dispersion of the electronic excitation. 
To include inner-shell electronic excitations in the computation of the ELF we add Drude-Lorentz functions multiplied by a switching function $F$, in order to reproduce the sharp edges of core excitations.

\begin{eqnarray}\label{Im}
 \textrm{Im}\Big[ \frac{-1}{\bar{\epsilon}({\bf{q}=0},W)}\Big]_{\rm}= \nonumber\\ 
 =\sum_{i} \frac{A_i \gamma_i W}  {(W_i^2-W^2)^2+(\gamma_i  W)^2} \cdot F(W,B_i) 
\end{eqnarray}
where:

\begin{equation}\label{F}
     F(W,B_i)=
    \left\{\begin{array}{lr}
     1 \textrm{ for outer electrons} \\
      \frac{1}{1+e^{-(W-B_i)}} \textrm{ for inner electrons} 
    \end{array}\right.
  \end{equation}
In Eqs. \ref{Im}-\ref{F} $A_i$, $W_i$, and $\gamma_i$, $B_i$ are fitting parameters. Finally, the momentum dispersion is introduced by \cite{Ritchie1977}:

\begin{eqnarray}\label{disp}
 W_i(q) = \sqrt{{W_i}^2+(12/5) \cdot \mathrm{E_f} \cdot  q^2/2+q^4/4} \nonumber \\
 \gamma_i(q) = \sqrt{{\gamma_i}^2+ q^2/2+q^4/4}
\end{eqnarray}
where $E_f$ is the Fermi energy. The parameters $B_i(q)$ have the same dispersion of $W_i(q)$ for consistency.

\begin{figure}[htpb!]
\centering
\includegraphics[width=0.5\textwidth]{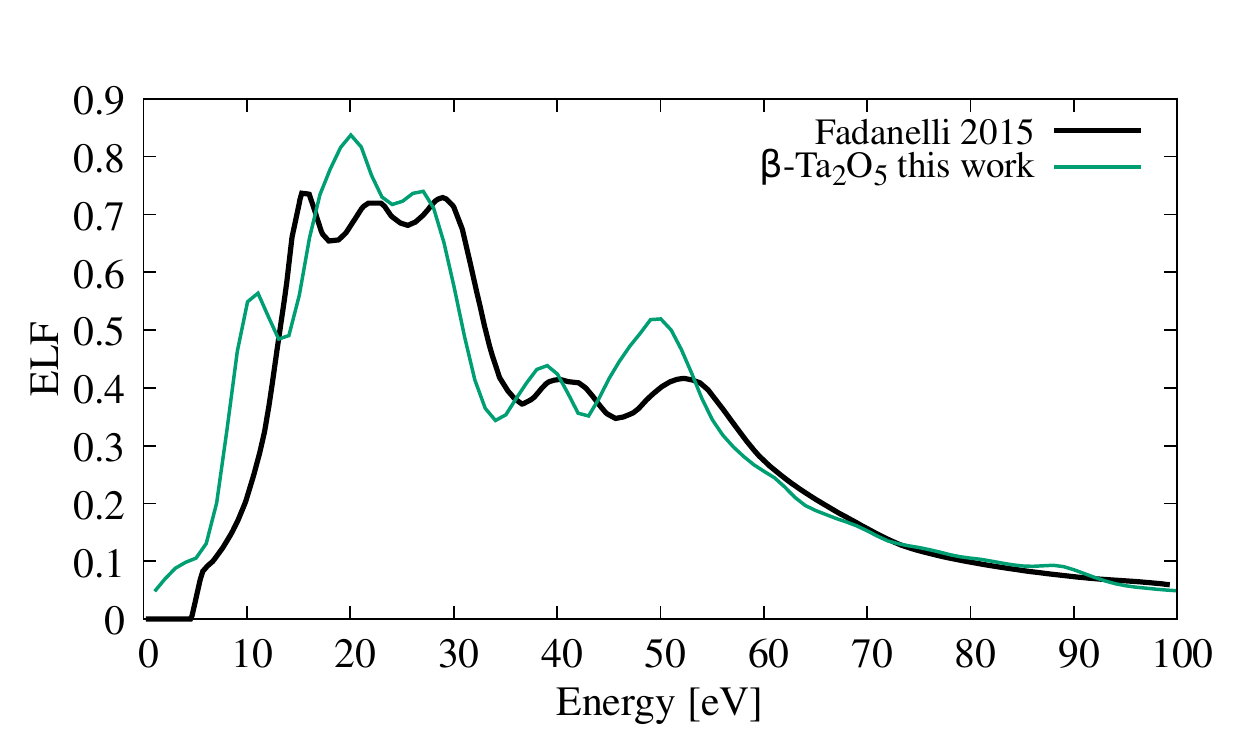}
\includegraphics[width=0.5\textwidth]{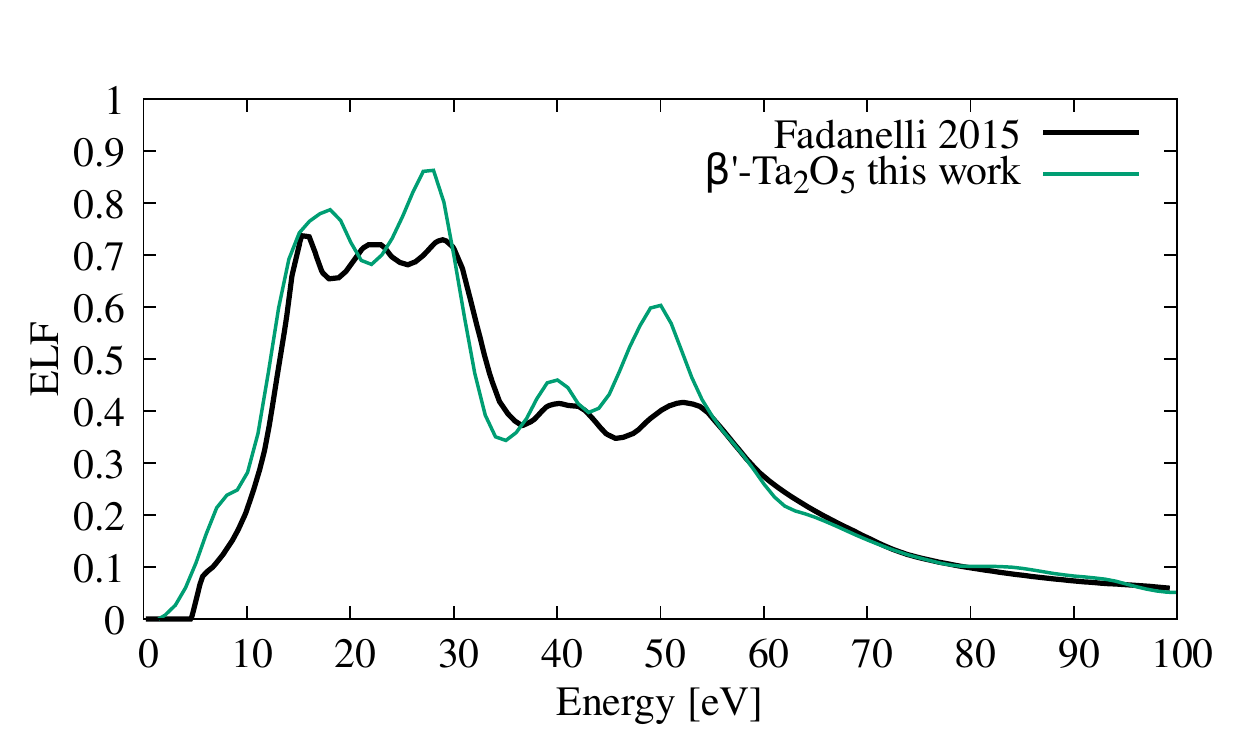}
\includegraphics[width=0.5\textwidth]{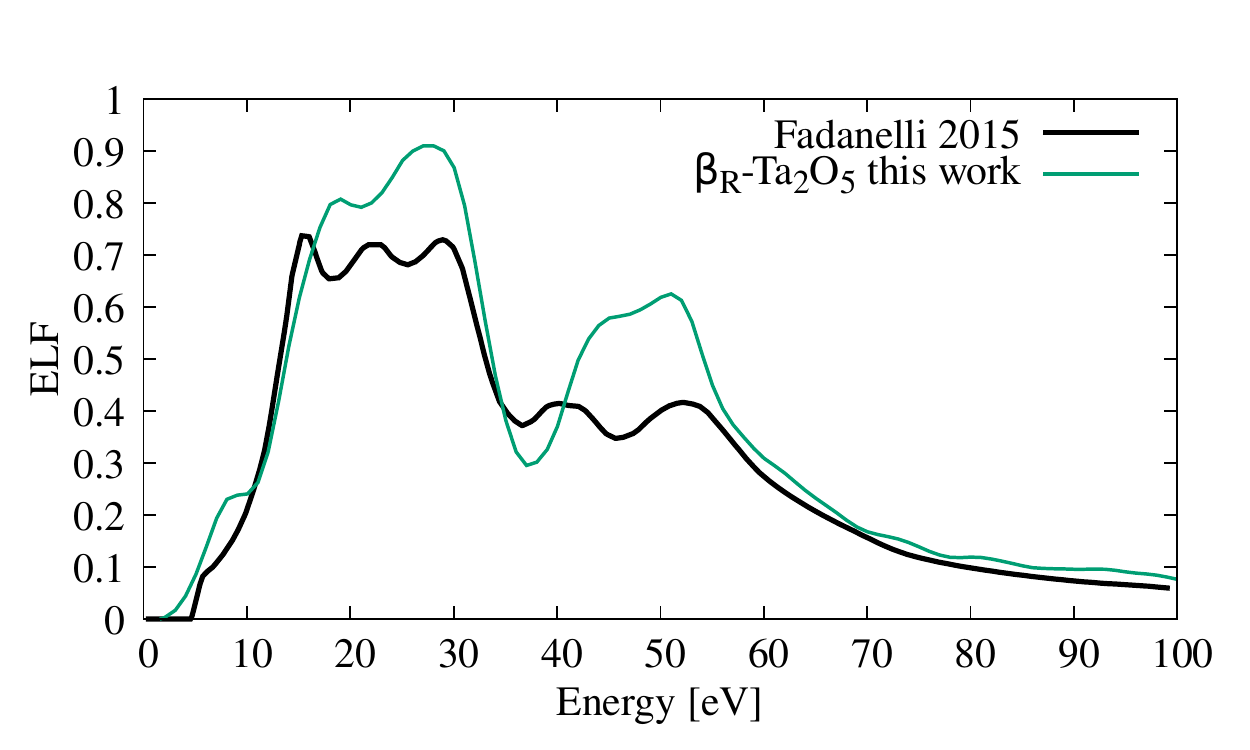}
\caption{ELF of $\beta$-Ta$_2$O$_5$, $\beta'$-Ta$_2$O$_5$, $\beta_{\mathrm{R}}$-Ta$_2$O$_5$ polymorphs in comparison with experimental data from Ref. \cite{Fadanelli2015}.}
\label{fig:ELFPOLI1}
\end{figure}

\begin{figure}[htpb!]
\centering
\includegraphics[width=0.5\textwidth]{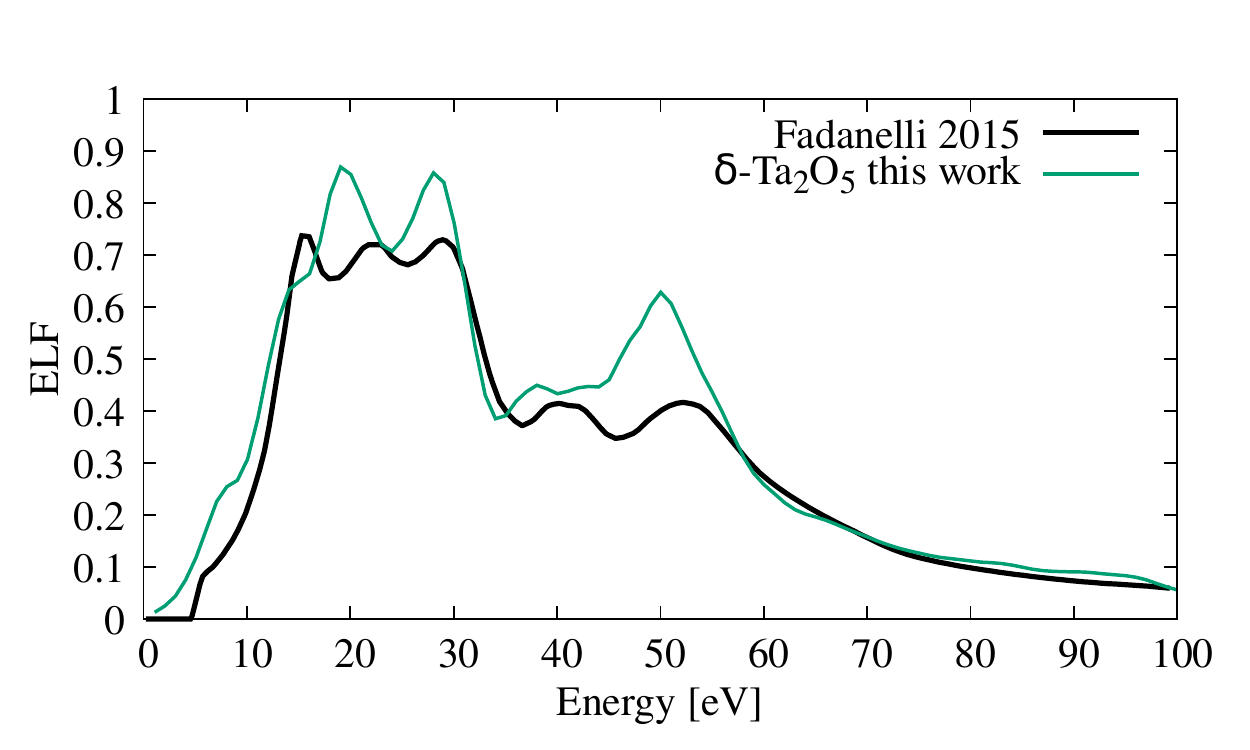}
\includegraphics[width=0.5\textwidth]{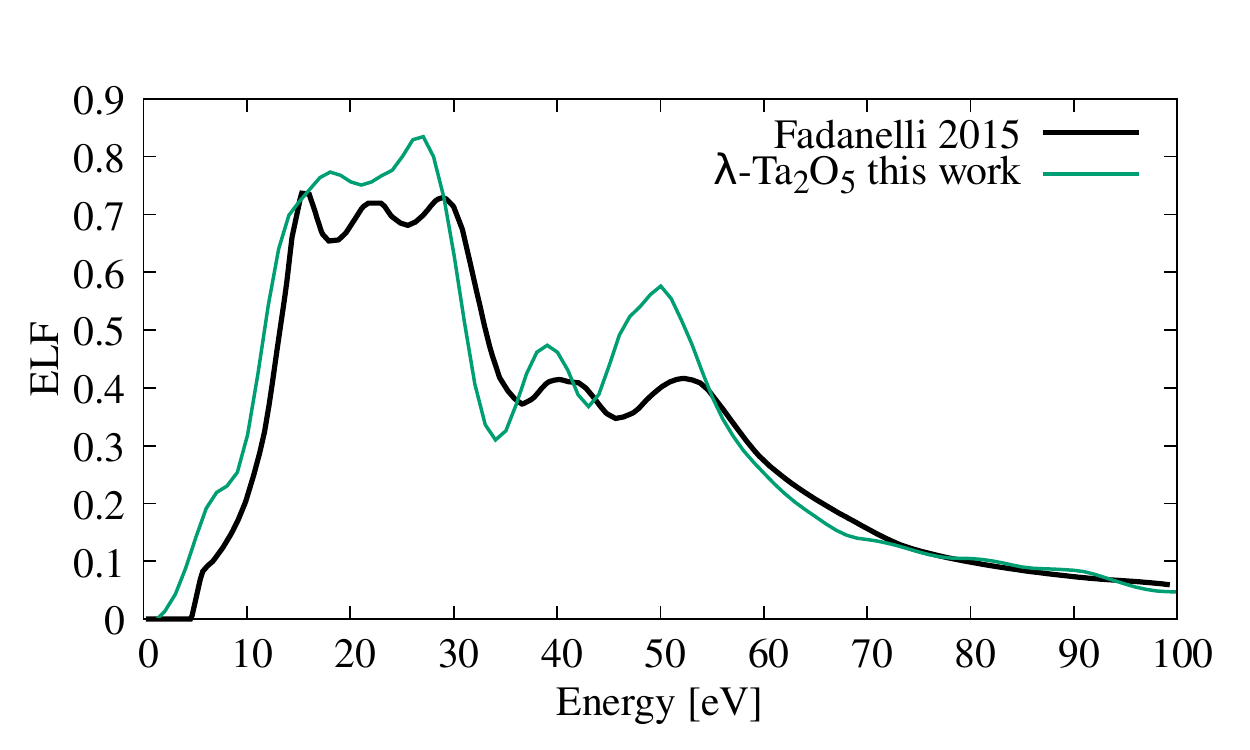}
\caption{ELF of $\delta$-Ta$_2$O$_5$ and $\lambda$-Ta$_2$O$_5$ polymorphs in comparison with experimental data from Ref. \cite{Fadanelli2015}.}
\label{fig:ELFPOLI2}
\end{figure}

\subsection{Monte Carlo simulations of electron transport}

With the ELF extension in place, Monte Carlo inputs, such as the inelastic scattering cross section, can be obtained. In the MC method for charge transport electrons are considered as point particles following classical trajectories that are induced by elastic and inelastic interactions. However, the elastic and inelastic scattering processes are dealt with quantum mechanics.
The occurrence probabilities of elastic and inelastic interactions at a certain kinetic energy $T$ can be calculated, respectively, by $p_\mathrm{el} (T)={\Lambda_\mathrm{el}} (T)/{\Lambda_{\mathrm{tot}}} (T)$ and $p_\mathrm{inel} (T)= {\Lambda_\mathrm{inel}} (T)/{\Lambda_{\mathrm{tot}}}(T)$, where $\Lambda_\mathrm{tot}(T) = \Lambda_\mathrm{inel}(T) + \Lambda_\mathrm{el}(T)$
and
$\Lambda_{\rm el/inel}=\lambda_{\rm el/inel}^{-1}(T)$ is the inverse mean free path for the elastic or inelastic scattering \cite{Dapor2020book}. The Monte Carlo proceeds by selecting a particular interaction event by comparing a random number uniformly distributed in the range $[0,1]$ with the relevant probability.

In particular, the energy loss $W$ suffered by electrons with kinetic energy $T$ upon inelastic scattering events can be reckoned by equating the inelastic scattering cumulative probability distribution to a uniformly-distributed random number in the range $[0, 1]$, as follows:
\begin{equation}
\label{inelcum}
P_\mathrm{inel}(T,W) = \frac{1}{\Lambda_\mathrm{inel}(T)} \int_{0}^{W} \frac{d \Lambda_\mathrm{inel}(T,W')}{dW'}dW' 
\end{equation}
Other than energy loss, the angular deviation due to inelastic scattering events is evaluated according to the classical binary collision theory \cite{Dapor2020book}.
In Eq. \ref{inelcum} the integrand is the differential inverse inelastic mean free 
path (DIIMFP):
\begin{equation}
\frac{d\Lambda_\mathrm{inel}(T,W)}{dW} = \frac{1}{\pi T} \int_{q_-}^{q_+} \frac{1}{q} \textrm{Im} \left [ \frac{-1}{\bar{\epsilon}(\mathbf{q}, W)} \right ]dq, 
\label{EqDIIMFP}
\end{equation}
where the integration limits
$q_{\pm} = \sqrt{2T} \pm  \sqrt{2(T-W)}$,
result from momentum conservation during the interaction process.

From the DIIMFP the inelastic mean free path (IMFP) can be derived. The latter is defined as the inverse of the inelastic scattering cross section:
\begin{equation}
\Lambda_\mathrm{inel}(T) = \int_{W_\mathrm{min}}^{W_\mathrm{max}} \frac{d\Lambda_\mathrm{inel}(T,W)}{dW} dW,
\label{EqIMFP}
\end{equation}
where the integration limit $W_\mathrm{min}$ is set to $E_\mathrm{gap}$ for semiconductors and insulating materials, while $W_\mathrm{max}$ represents the minimum between $T$ and $(T+W_\mathrm{min})/2$. 
Dealing with a bulk system, surface plasmon excitations are not taken into account. 

The second key quantity to run Monte Carlo simulations of charge transport in solids is the elastic scattering cumulative probability. In this respect, the angular deviation of the electron trajectories following an elastic collision in terms of the scattering angle $\theta$ is 
\begin{equation}
    \label{elcum}
    P_\mathrm{el}(T, {{\theta}}) = \frac{2\pi}{\Lambda_\mathrm{el}(T)} \int_0^{{\theta}}  \frac{d\Lambda_\mathrm{el}(T,\theta')}{d\Omega} \sin\theta' d\theta'.
\end{equation} 
$\theta$ can be evaluated by equating this cumulative probability to a uniformly distributed random number generated in the range $[0,1]$. Electrons do not suffer energy loss during an elastic collision.

The elastic scattering is accounted for by means of the Mott theory \cite{Mott1929}. This approach provides the differential elastic scattering cross-section (DESCS) for an electron impinging on a central potential and subsequently scattered by an angle $\theta$ as follows \cite{Dapor2020book}:
\begin{equation}\label{molsca}
\frac{d\Lambda_{\mathrm{el}}(T,\theta)}{d\Omega}\,={\cal N}\,\sum_{m,n}\,e^{(i {\bf q} \cdot {\bf r}_{mn})}\,[f_m(\theta) f^*_n(\theta)\,+\,g_m(\theta) g^*_n(\theta)]\,,
\end{equation}
where ${\bf r}_{mn}\,=\,{\bf r}_m\,-\,{\bf r}_n$, and ${\bf r}_m~({\bf r}_n)$ identifies the position of the $m^{\mathrm{th}}~(n^{\mathrm{th}})$ atom in the periodic unit cell, ${\cal N}$ is the target atomic number density, $f_m(\theta)$ and $g_m(\theta)$ are the direct and spin-flip scattering amplitudes of the $m^{\mathrm{th}}$--atom and of the neighbouring $n^{\mathrm{th}}$--atom, respectively,  which can be obtained by solving the Dirac equation in a central atomic field. Eq. \ref{molsca} generalizes the Mott theory for the scattering from atomic targets by taking into account the presence of bonded interactions among neighbours in the periodic unit cell. Indeed, by allowing interference between the direct and spin-flip  scattering amplitudes condensed phase effects can be included in the assessment of the elastic scattering cross section. 

Finally, by integrating over the solid angle one obtains the total elastic scattering cross section:
\begin{equation}
 \Lambda_{\mathrm{el}}(T) = \int_{\Omega} \frac{d \Lambda_{\mathrm{el}}(T,\theta)}{d \Omega} {d \Omega}.
\end{equation}

The calculation of the DESCS has been performed by the ELSEPA code \cite{Salvat2005} using a 2 $\times$ 2 $\times$ 1 supercell to account for crystalline effects over the elastic scattering cross section.

The MC calculation has been performed using the SEED code \cite{Dapor2020book}. The electron trajectories ensemble is set to reach statistical significance and a low noise to signal ratio of the simulated data ($\approx 10^8$ trajectories). The full width at half maximum (FWHM) of the elastic peak has been set to $0.8$ eV in order to match the experimental one. 

\section{Results and Discussion}

\subsection{Energy loss function from ab initio calculations}

As many different structural models were reported for Ta$_2$O$_5$, we investigate here the ELF of the most relevant geometries. This quantity is indeed a relevant marker that can be compared with available experimental data for the identification of the oxide atomic structure. 

\begin{figure}[htbp!]
\centering
\includegraphics[width=0.5\textwidth]{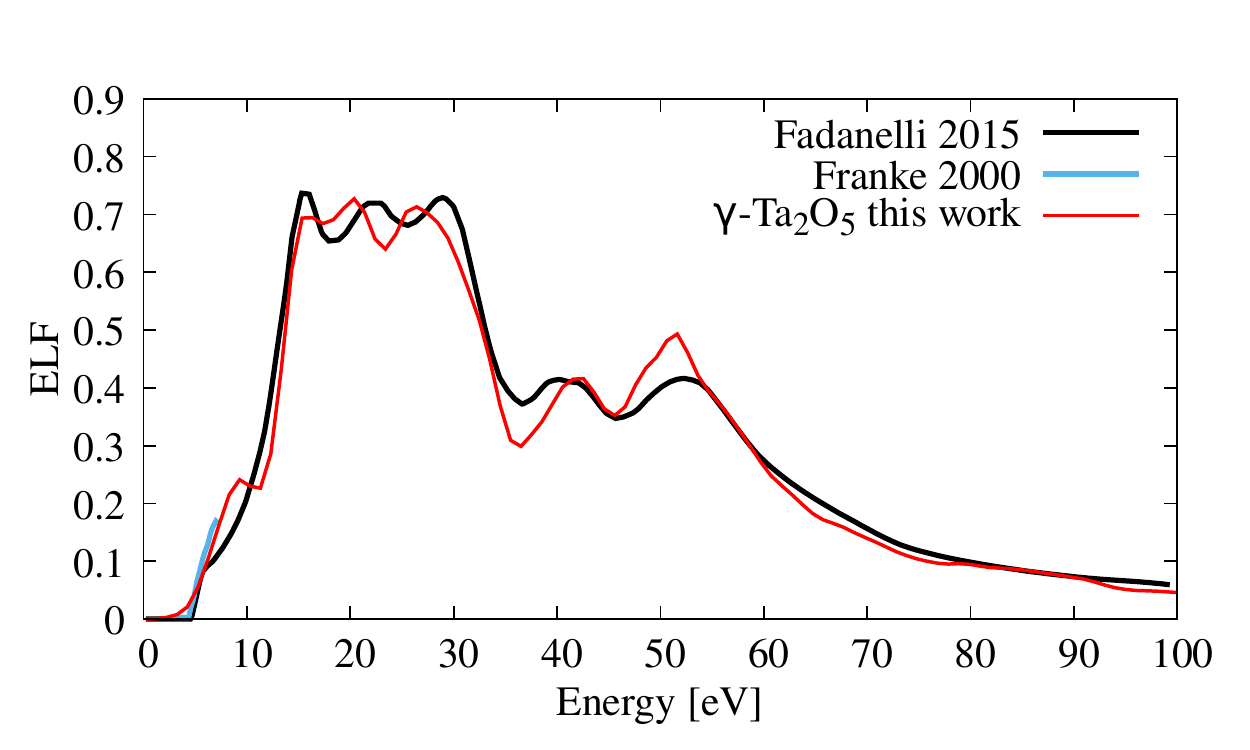}
\includegraphics[width=0.5\textwidth]{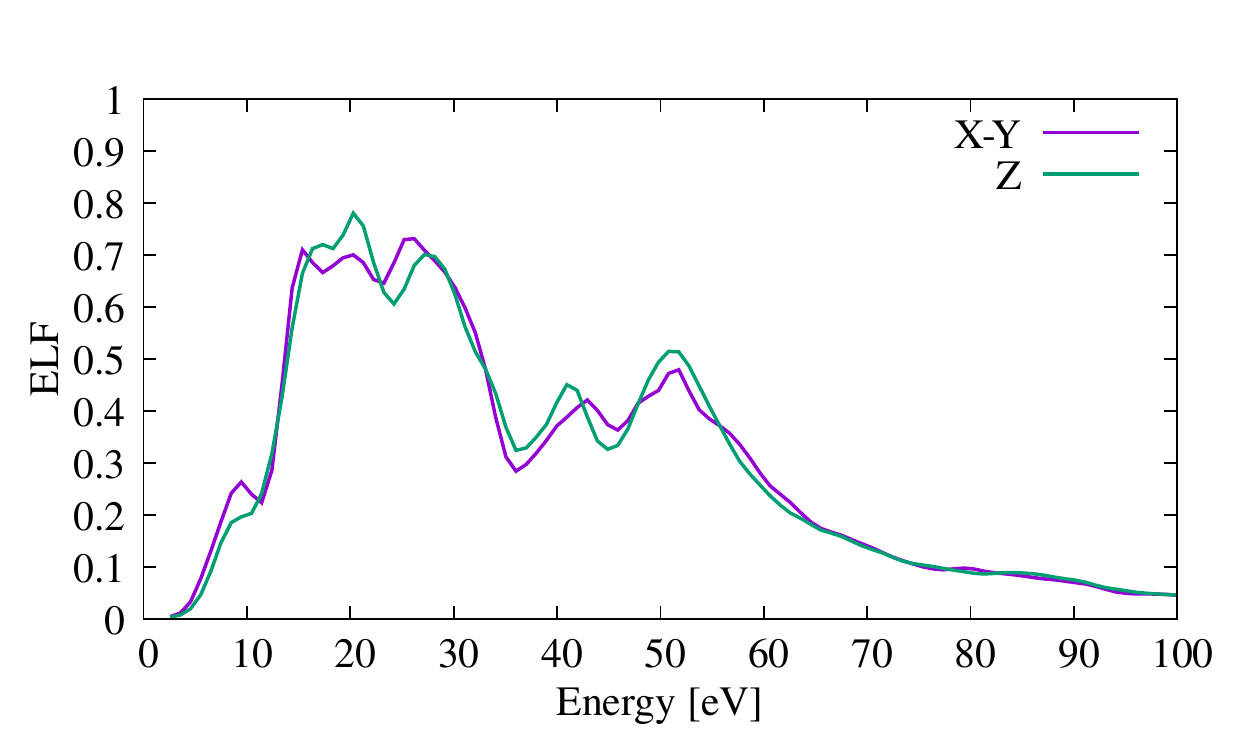}
\includegraphics[width=0.5\textwidth]{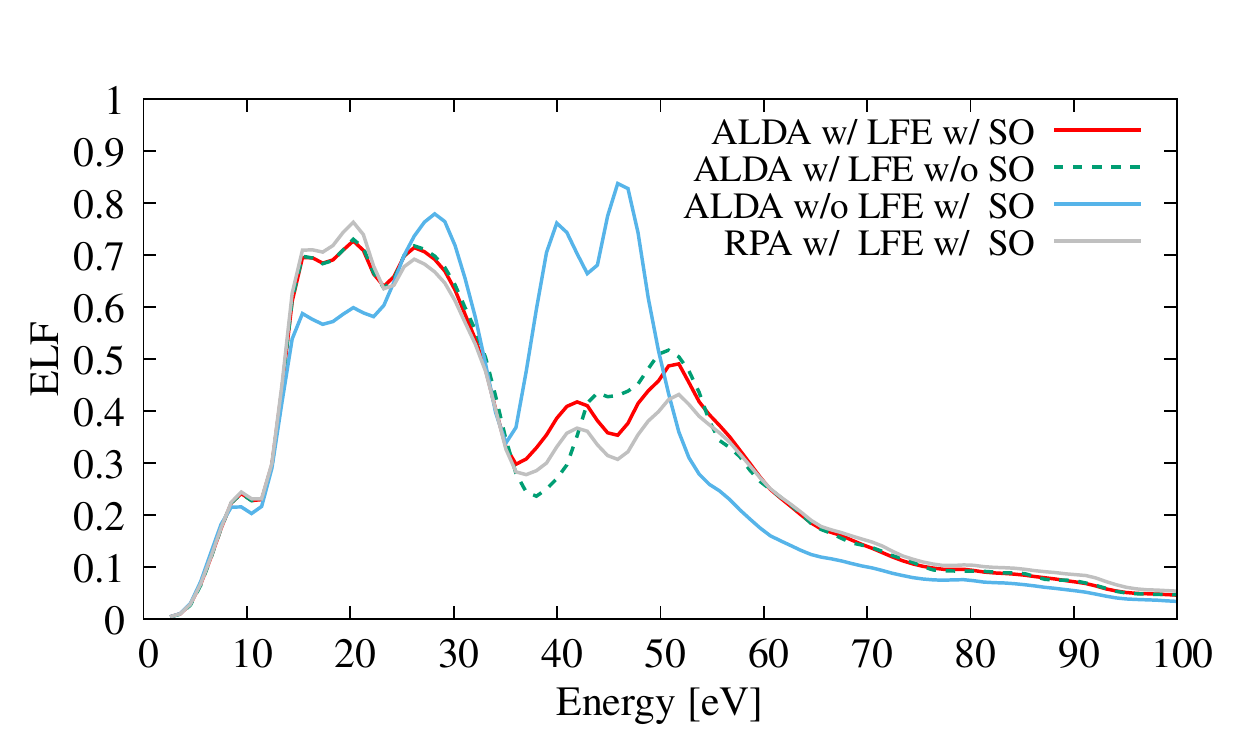}
\caption{Top panel: ELF of $\gamma$-Ta$_2$O$_5$ in comparison with experimental data from Refs. \cite{Fadanelli2015, Franke2000}. Central panel: Decomposition of the ELF along the polarization directions ($X-Y,Z$). Bottom panel: ELF at different levels of theory. Calculated ELF are blueshifted by $1.5$ eV.}
\label{fig:ELFGAMMA}
\end{figure}

We report in Figs. \ref{fig:ELFPOLI1} and \ref{fig:ELFPOLI2} the comparison of our calculated ELFs of several polymorphs with those that can be obtained from the REEL rough experimental data recorded on a Ta$_2$O$_5$ sample by reverse quantitative analyses of the electron energy loss spectra using the QUEELS code \cite{Fadanelli2015}. We notice that the sample under investigation \cite{Fadanelli2015} is a $50$ nm thick layer of Ta$_2$O$_5$ that was grown on a Ta substrate by thermal oxidation at $600$ °C upon oxygen irradiation. The ELF of the $\gamma$-Ta$_2$O$_5$ phase is reported in Fig. \ref{fig:ELFGAMMA}, where a rigid blueshift of $1.5$ eV (used in all theoretical spectra) has been applied to our calculated data to align them with the results refined using the QUEELS software package from the experimental EELS. This shift can be rationalised so as to correct the well-known underestimation of the band gap, which impacts the ELF calculations via Eq. \ref{EqIMFP}, using the LDA ($2.98$ eV in this work) and PBE functionals ($2.26$ eV \cite{Yang2018}) with respect to higher accuracy approaches, such as the HSE06 ($3.75$ eV) and PEB0 ($4.51$ eV) hybrid functionals \cite{Yang2018}. From absorbance measurements, values ranging from 4.2 to 4.4 eV for amorphous films and 3.9 to 4.5 eV for crystalline films were obtained \cite{CHANELIERE1998269}.

We notice in the top panel of Fig. \ref{fig:ELFPOLI1} that the spectral lineshape derived from experiments presents five well resolved peaks. In particular, the first three peaks in the range $15-35$ eV are characterised by a comparable intensity, while those at high energy ($>35$ eV) have lower intensity. These characteristics are also reproduced in our calculations (see green lines in Figs. \ref{fig:ELFPOLI1} and \ref{fig:ELFPOLI2}). However, the spectral lineshape of these polymorphs cannot be considered in good agreement with the reference ELF (black line in Figs. \ref{fig:ELFPOLI1} and \ref{fig:ELFPOLI2}). In fact, in the range $15-35$ eV none of these structural arrangements deliver the three peaks with similar intensity in the experiments. Furthermore, only $\beta$-Ta$_2$O$_5$, $\beta'$-Ta$_2$O$_5$,  $\lambda$-Ta$_2$O$_5$ polymorphs are characterised by two well resolved peaks at higher energy, which are at variance merged in one broad peak for the $\beta_{\mathrm{R}}$-Ta$_2$O$_5$
phase, while the first of them is almost suppressed in the $\delta$-Ta$_2$O$_5$ form.

At odds with these findings, a remarkable agreement with the experimental-derived data can be obtained by considering the $\gamma$-Ta$_2$O$_5$ phase, as shown in Fig. \ref{fig:ELFGAMMA}. In particular, the three peaks of similar intensity in the range $15-35$ eV are recovered, and the two peaks beyond $35$ eV are less bright than for the other cases, in better agreement with the ELF experimental data. Moreover, the shoulder at $10$ eV, while not so pronounced in the spectra from Ref. \cite{Fadanelli2015}, is clearly reproduced in our simulations as found in Ref. \cite{Franke2000}. We stress that while the ELF reported in Ref. \cite{Fadanelli2015}, which is derived from the REELS experiments by using the QUEELS code, has been corrected at energy larger than $ 60$ eV to fulfil the sum rules, our ab initio calculations do not need any correction.
Finally, from the analysis of the spectra presented in Figs. \ref{fig:ELFPOLI1}, \ref{fig:ELFPOLI2}, and \ref{fig:ELFGAMMA} we conclude that the ELF lineshape of the $\gamma$-Ta$_2$O$_5$ phase is the only to follow tightly the reference experimental data \cite{Fadanelli2015}, which makes it the most suitable candidate as model structure of pentoxides. 

Furthermore, the contribution to the ELF coming from different polarization directions, is reported in Fig. \ref{fig:ELFGAMMA} (middle panel). We note that the main features of the ELF are essentially independent on the polarization direction aside from the shoulder at $10$ eV and the central peak among the triplets in the range $15-35$ eV.

To show the significant impact that the inclusion of both the LFE and the SO coupling has on the calculated lineshape, in the bottom panel of Fig. \ref{fig:ELFGAMMA} we report the ELF of the $\gamma$-Ta$_2$O$_5$ phase with/without the LFE and the SO coupling, respectively, as well as in the random phase approximation (RPA). 
We stress that the role of LFE is relevant: they affect all the features of the ELF demonstrating the presence of strong spatial inhomogeneity. 

In particular, the LFEs mainly decrease the intensity of the peaks in the range $35-60$ eV, showing their fundamental role in reproducing the experimental intensity of the three peaks, while they have a smaller effect on the low energy part of the spectrum.
The SO coupling has a less relevant impact on the ELF with respect to the LFE, but it reveals crucial in order to recover the peak at $35-45$ eV.
We also notice that the difference in using the ALDA kernel or RPA is essentially negligible both with and without introducing the LFEs (not shown).

\subsection{Energy extension and momentum dispersion of the ELF}

\begin{figure}[htbp!]
\centering
\includegraphics[width=0.5\textwidth]{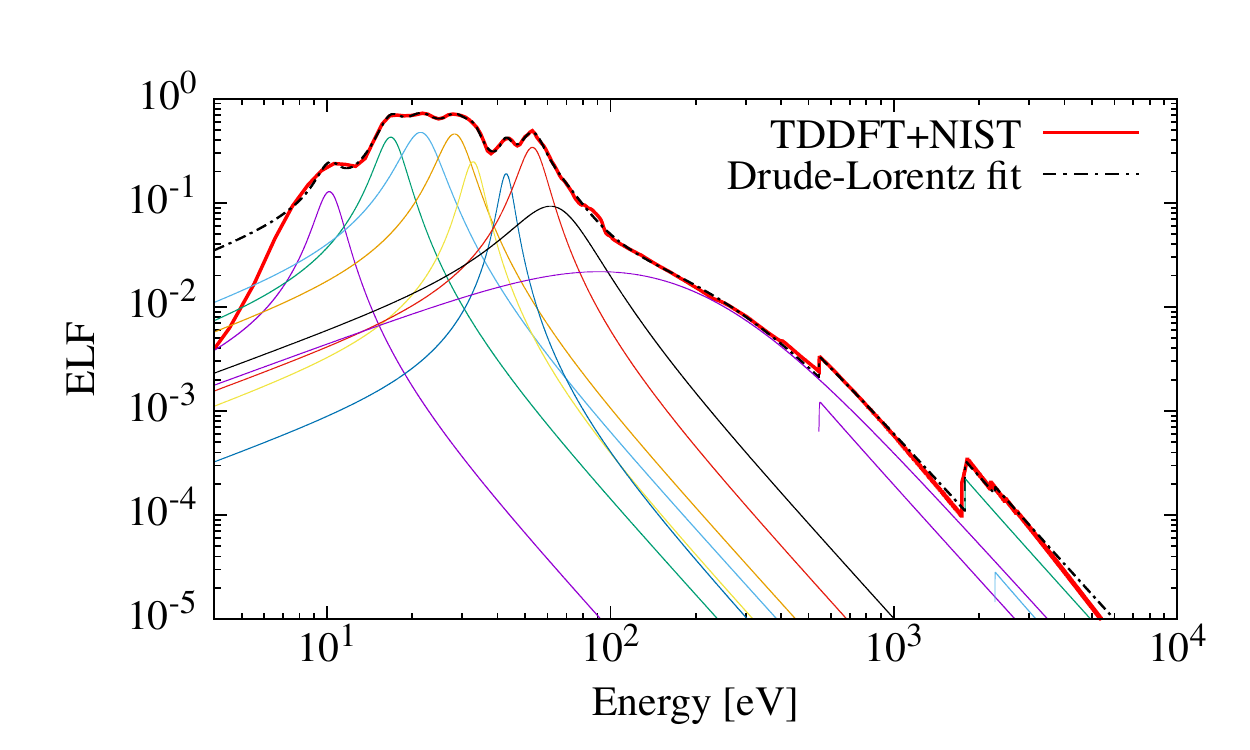}
\includegraphics[width=0.5\textwidth]{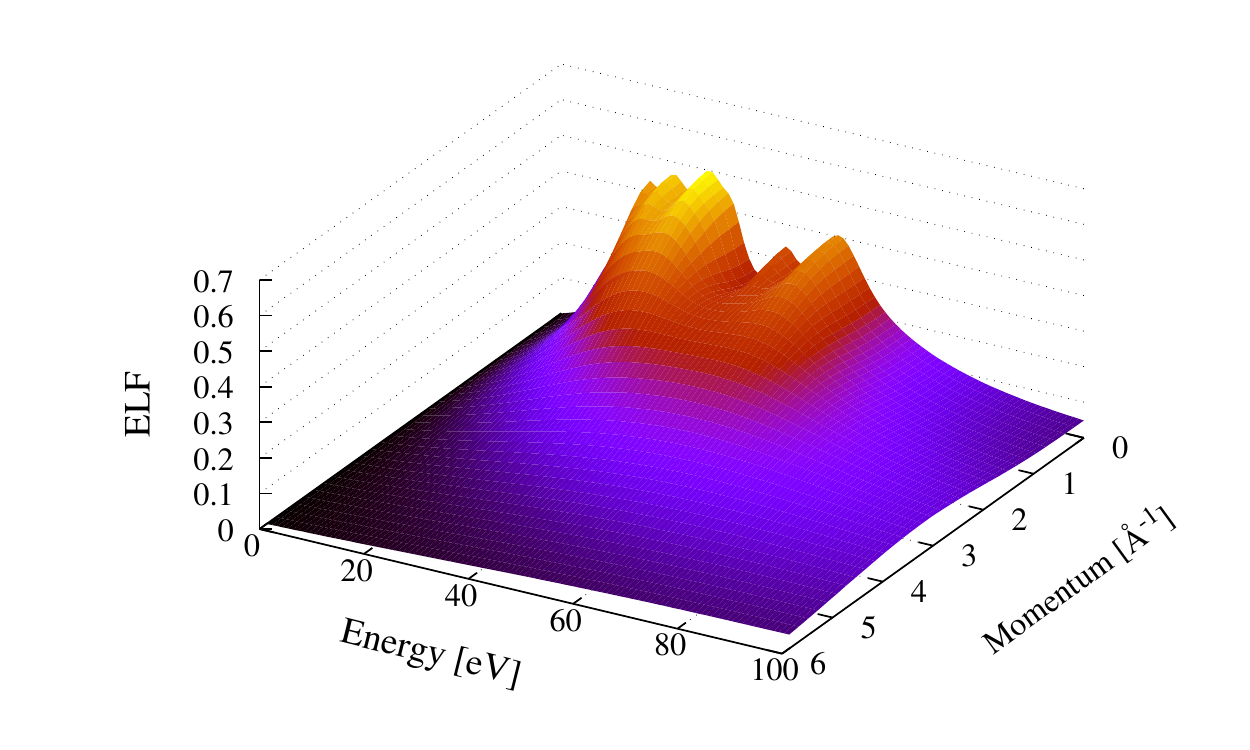}
\caption{Top panel: ELF of the $\gamma$-Ta$_2$O$_5$ form in the optical limit and the relevant fitting functions. Bottom panel: Bethe surface of $\gamma$-Ta$_2$O$_5$ as a function of transferred momentum and excitation energy by extending to finite momentum the optical ELF via a momentum dependent Drude--Lorentz model.}
\label{fig:ELFfit}
\end{figure}

To run MC simulations one needs to determine the dependence of the ELF on the entire spectrum of excitation energies and its dispersion with respect to momentum transfer (the so-called Bethe surface). 
To achieve this goal we used the Drude-Lorentz model with a momentum dependent broadening discussed in Eqs. \ref{Im}-\ref{disp}. 
In Fig. \ref{fig:ELFfit} we report the $\log-\log$ plot of the optical ELF and the fitting Drude functions. The fit parameters $A_i$, $W_i$, $\gamma_i$, and $B_i$ are reported in Tab. \ref{tab:Parameters}

\begin{table}[b]
\caption{\label{tab:Parameters} Parameters $A_i$, $W_i$, $\gamma_i$, and $B_i$ of the Drude-Lorentz fit of the optical ELF.}
\begin{center}
\begin{tabular}{@{}cccc}
\br
$A_i$ (eV$^2$)	&	$W_i$ (eV)	&	$\gamma_i$ (eV) & $B_i$ (eV)\\
\mr
3.2	&	10.3	&	2.4	&\\
31.2	&	16.9	&	4.3&\\

76.8	&	21.7	&	7.5	&\\

108.8	&	28.5	&	8.4	&\\

50.8	&	32.9	&	6.2&	\\

47.1	&	43.0	&	5.8	&\\

238.7	&	53.4	&	13.1&	\\

242.4	&	64.8	&	41.4&	\\

1094.2	&	176.3	&	391.5	\\

1545.3	&	127.6	&	123.8 & 544.6	\\

11509.9	&	306.9	&	104.4 & 1783.4	\\

3000.5	&	482.8	&	101.8 & 2274.0	\\
\br
\end{tabular}
\end{center}
\end{table}

\begin{figure}[htp!]
\centering
\includegraphics[width=0.5\textwidth]{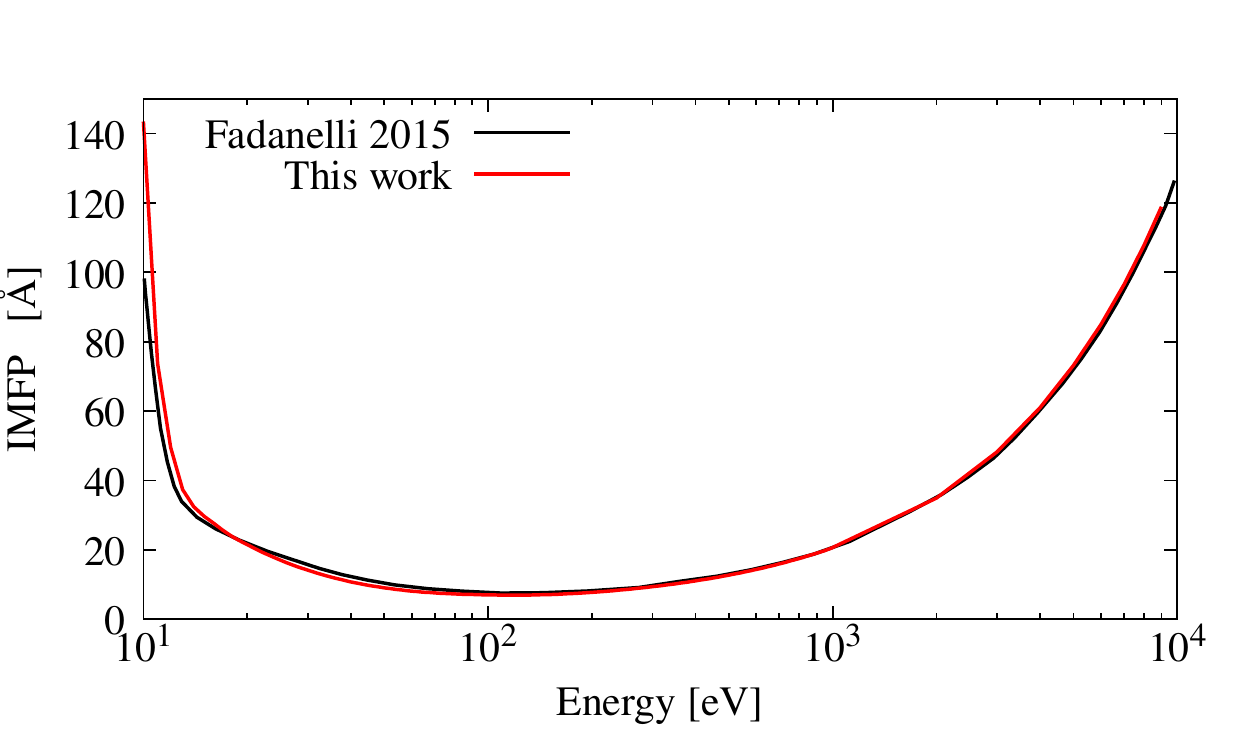}
\includegraphics[width=0.5\textwidth]{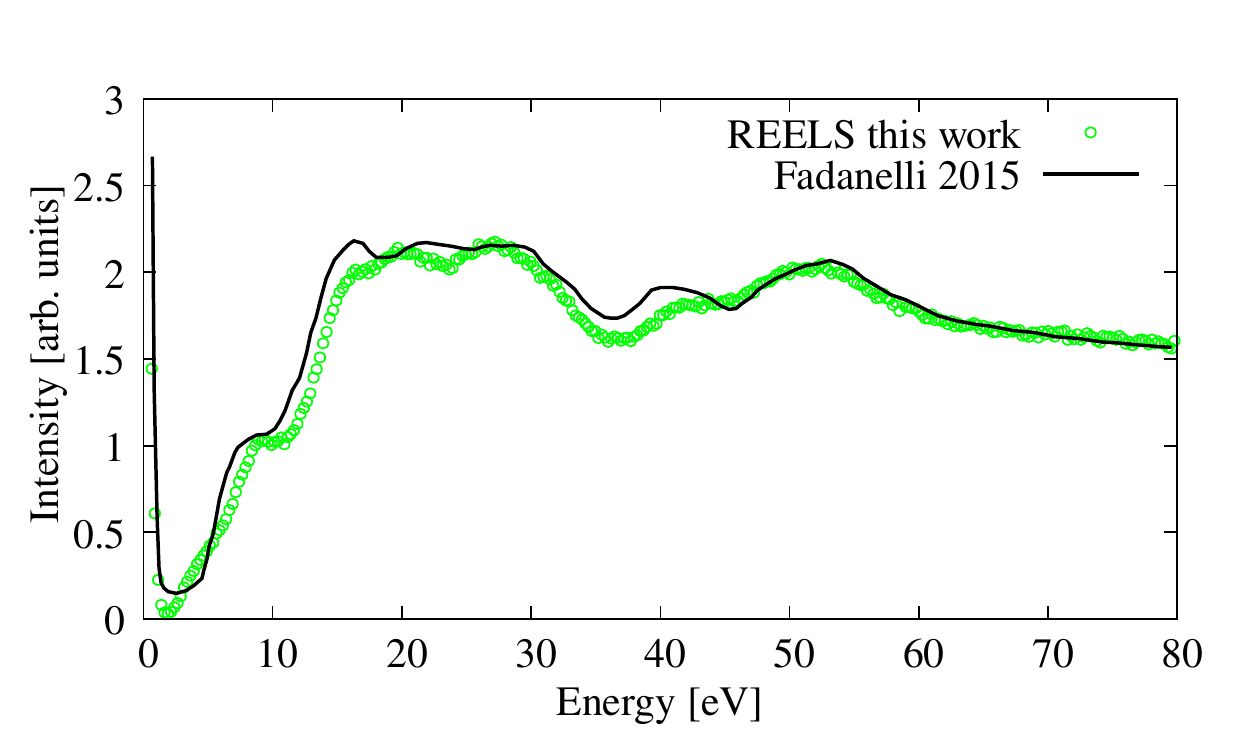}
\caption{Top panel: Comparison of our calculation of the IMFP with data from Ref. \cite{Fadanelli2015}. Bottom panel: Comparison of our Monte Carlo calculation of the REEL spectrum with experimental data from Ref. \cite{Fadanelli2015}.}
\label{fig:IMFP}
\end{figure}

A 3D plot of the momentum dispersion of the ELF related to the outer shell excitations is reported in Fig. \ref{fig:ELFfit}.
While we chose a specific analytical momentum dependence of the ELF (see Eq. \ref{disp}), we can consider our findings robust with respect to different dispersion models \cite{azzolini2017monte,azzolini2020comparison}, also in view of the relatively high kinetic energy of the impinging electron beam (5 keV). This can be seen for example by computing the IMFP of bulk Ta$_2$O$_5$ (see Eq. \ref{EqIMFP}), which we report in the top panel of Fig. \ref{fig:IMFP} alongside the data derived from the experiments \cite{Fadanelli2015}. We note that the calculated IMFP is in excellent agreement with the experimentally derived one \cite{Fadanelli2015} at all energies, although the dispersion relations used are different.

\subsection{Monte Carlo simulations of REEL spectra}

To compare directly the theoretical spectrum with the experimental energy loss lineshape of Ta$_2$O$_5$ \cite{Fadanelli2015}, rather than the ELF, which is derived from the former via a subtraction procedure \cite{Tougaard2004}, we  used Monte Carlo (MC) simulations. 
Indeed, to further validate the $\gamma$-Ta$_2$O$_5$ form as a good structural model for thermal tantalum oxide we decided to follow the inverse route and compute the REEL spectrum starting form our ab-initio ELF to be compared with the experimental measurements, in this way effectively avoiding any subtraction.

In the lower panel of Fig. \ref{fig:IMFP} we report the computed REEL spectrum of bulk  $\gamma$-Ta$_2$O$_5$ (solid green line) in comparison to the as-measured experimental data \cite{Fadanelli2015} (black line), finding an excellent agreement. We indeed reproduce all the main features of the experimental spectrum. The spectra were normalized with respect to the highest energy loss peak.

We stress that the very good agreement of our results with experimental REELS on the one hand validates our methodological approach, on the other hand confirms that the $\gamma$-Ta$_2$O$_5$ polymorph is the most suitable model of tantalum oxide grown by thermal oxidation. 

\section{Conclusions}

In summary, we computed the optical ELF of several polymorphs of Ta$_2$O$_5$ using a high-accuracy first-principles linear response TDDFT approach. We showed that the recently proposed $\gamma$-Ta$_2$O$_5$ polymorph is the most reliable structure in reproducing the ELF experimental lineshape. Furthermore, we have shown that the inclusion of LFE and SO coupling is fundamental for the description of the electronic structure of these material phases. In order to test our calculations directly against the experimental measurements, avoiding any subtraction, we used our ab initio optical ELF and the Drude--Lorentz approach for including the finite transfer momentum dispersion in the main quantities necessary to carry out MC calculations, such as the inelastic scattering cross section at different energies. We performed MC simulations of the REEL spectrum of $\gamma$-Ta$_2$O$_5$ polymorph finding an excellent agreement with the experimental data. These results represent a striking confirmation that the $\gamma$-Ta$_2$O$_5$ polymorph is the most suitable model for tantalum oxide grown by thermal oxidation. Our results can be useful to a further investigation of the dielectric properties of tantalum oxides, paving the way towards the application of these materials e.g. as radioenhancers in hadrontherapy for cancer cure.





\section*{Acknowledgments}

A. P. acknowledges Fondazione Caritro and DICAM (University
of Trento) for the financial support under the CARITRO project
High-Z ceramic oxide nanosystems for mediated proton cancer
therapy.

\section*{References}

\providecommand{\newblock}{}

\end{document}